\documentclass[twocolumn,preprintnumbers,amsmath,amssymb,aps,prl,longbibliography]{revtex4-1}
\usepackage{graphicx}
\begin{document}

\title{Colloidal Dynamics on a Choreographic Time Crystal}

\author{Andr\'as Lib\'al$^{1,2}$,  T\"unde Bal\'azs$^{2}$,
 C. Reichhardt$^{1}$ and C. J. O. Reichhardt$^{1}$}
\affiliation{$^{1}$Theoretical Division and Center for Nonlinear Studies,
Los Alamos National Laboratory, Los Alamos, New Mexico 87545, USA}
\affiliation{$^{2}$Mathematics and Computer Science Department, Babe{\c s}-Bolyai University, Cluj, 400084, Romania }

\date{\today}

\begin{abstract}
A choreographic time crystal is a dynamic lattice structure in which the points comprising the lattice move in a coordinated fashion. These structures  were initially proposed  for understanding the motion of synchronized satellite swarms.  Using simulations we examine colloids interacting with a choreographic crystal consisting of traps that could be created optically. As a function of the trap strength, speed, and colloidal filling fraction, we identify a series of phases including states where the colloids organize into a dynamic chiral loop lattice as well as a frustrated induced liquid state and a choreographic lattice state.  We show that transitions between these states can be understood in terms of vertex frustration effects that occur during a certain portion of the choreographic cycle.  Our results can be generalized to a broader class of systems of particles coupled to choreographic structures, such as vortices, ions, cold atoms, and soft matter systems.  
\end{abstract}

\maketitle

\vskip 2pc

Crystalline states arise throughout nature and are characterized 
by their symmetries.
Since these structures are static in time, they can be described by
a single snapshot.
Recently there
have been proposals for dynamic crystals containing points that
move in a synchronized fashion
such that a single time snapshot
does not reveal all the symmetries in the system.
These structures are called choreographic crystals \cite{Boyle16},
and they are composed of
a collection of points
that undergo
a series of repeated moves
to form
varying patterns that recur over time.  
A related idea is that in some such systems, 
the ground states themselves
are also
periodic in time,
forming what are called time crystals \cite{Shapere12,Sacha18}.
Generally, time crystal systems must be driven
out of equilibrium 
and contain
some form of dissipation,
so they are not in a true ground state.
Nevertheless,
there is growing interest in creating 
and studying the properties of classical 
\cite{Shapere12,Sacha18,Flicker18,Yao18,Dai19}
and quantum \cite{Wilczek12,Li12,Zhang17} time and choreographic 
crystals in  condensed matter, atomic, and even cosmological systems
\cite{Boskovic18,Easson19}.   
Choreographic crystals represent a new type of structure
and there are many open questions, including
how to realize these states,
what their properties are, and whether they could be coupled to other
systems.

Here we examine a system of dynamic traps that form a choreographic crystal
coupled to an assembly of colloidal particles.
There have been many 
studies
of colloidal trapping on static crystalline substrates
\cite{Chowdhury85,Wei98,Korda02a,Reichhardt02a,Brunner02,Agra04,OrtizAmbriz16,Brazda18} 
or quasiperiodic lattices \cite{Mikhael08,Schmiedeberg08},
where melting and commensurate-incommensurate transitions were observed.
Studies of the dynamics of colloids driven over such crystalline substrates 
have revealed locking of the
motion of the colloids with a symmetry direction of the substrate lattice 
\cite{Korda02,MacDonald03,Reichhardt11,Bohlein12a,Cao19},
depinning of kinks and antikinks at incommensurate versus commensurate states
\cite{Bohlein12,Vanossi12,Hasnain13,McDermott13a},
and a diverse array of other dynamical phenomena \cite{Loehr16,Tierno07}.
It is even possible to dynamically control and move individual traps \cite{Grier03}
or to flash the traps on and off \cite{Libal06,Brazda17}, 
so with appropriate rules for translation,
it should be feasible to create a choreographic lattice of
optical traps that couple to colloidal particles.
Beyond colloids, optical trapping lattices
have been created for cold atom systems \cite{Buchler03,Muldoon12},
ions \cite{Schmidt18}, vortices in 
Bose-Einstein condensates \cite{Tung06},
and vortices in type-II superconductors \cite{Veshchunov16},
and therefore similar choreographic
lattices could be created for these systems.
Choreographic trap arrays thus represent a new type 
of lattice for studies of commensuration effects and dynamics. 

In our simulations, 
we find three generic phases of 
colloid dynamics depending on the strength and speed of the traps as well as the
filling fraction or ratio of the number of colloids to the number of traps.
In the weakly coupled regime, the colloids are temporarily trapped
and organize into a dynamical chiral loop crystal.
In the partially coupled regime, where a given colloid is dragged by a
trap for a varied length of time before decoupling from the trap,
a liquid like state appears.
In the strong coupling regime,
the colloids are permanently locked to the traps and themselves form
a choreographic crystal.
At  higher filling
fractions, we observe phases in which traps containing multiple colloids
interact with interstitial colloids in the regions between traps.
The transitions between the dynamical states
are affected by the trap velocity
since the colloids decouple from the traps at high trap velocities,
and
we map out a dynamic phase diagram as a function of trap strength and velocity.  
We also show that the transition into and out of the liquid phase
is the result of a vertex frustration effect, similar to
that found in triangular artificial spin ice \cite{Libal18,Nisoli18,OrtizAmbriz19},
which appears during the portion of the choreographic cycle
when the spacing between the traps reaches its minimum value.

\begin{figure}
\includegraphics[width=3.5in]{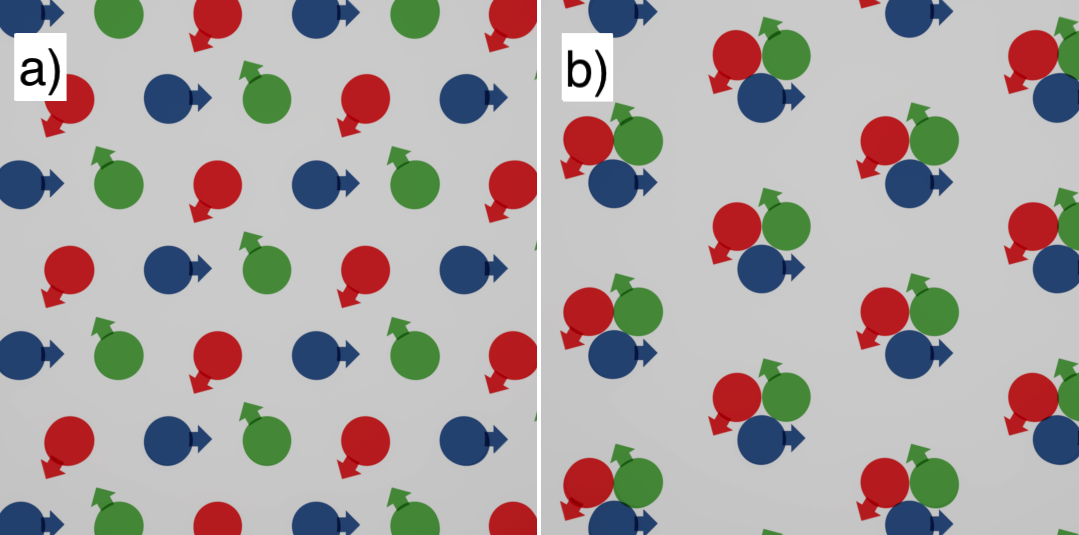}
\caption{(a) Schematic of a choreographic lattice composed of
  three subsets of traps (red, blue, and green) that move at a velocity
  $v_{\rm trap}$ in a synchronized manner according to the rules
  given in Ref.~\cite{Boyle16}.
  Each subset of traps moves in a different direction defined by
  the vector $(x,y)$ which has the value
  $(-0.5,-\sqrt{3}/2)$  for the red traps, $(1,0)$  for the blue traps,
  and $(-0.5,+\sqrt{3}/2)$ for the green traps, as
  indicated by the arrows.
  Under this motion, the traps never overlap, and they reform the original
  triangular lattice ordering shown in (a) every
  $\tau$ time units.
  (b) Image of the trap positions at the point in the cycle where the spacing
  between the traps reaches its smallest value.
}
\label{fig:1}
\end{figure}

{\it Simulation---}
We conduct simulations of point-like colloidal particles in a two dimensional box
of size $L \times L\sqrt{3}/2$ with $L=96.0$ where there are  
periodic boundary conditions in 
the $x$ and $y$ directions.
The sample contains
$N_{\rm trap}=576$ trapping sites
of radius $R_{\rm trap}=1.0$ which are initially arranged in a
$24\times 24$ hexagonal lattice with lattice constant $a = 4.0$,
large enough to ensure that traps never overlap
when they are translated. 
To create the choreographic crystal, we use
the rules for motion introduced in Ref.~\cite{Boyle16}.
The traps are divided into three subsets, as shown schematically in Fig.~\ref{fig:1}(a).
Each subset of traps moves in a direction given by the vector $(x,y)$, which has
the value
$(-0.5,-\sqrt{3}/2)$ for the first subset, $(1,0)$ for the second subset,
and $(-0.5,+\sqrt{3}/2)$ for the third subset.
The traps are initialized in a hexagonal lattice, as shown in Fig.~\ref{fig:1}(a), and each
trap moves in a straight line with a velocity $v_{\rm trap}$.
The original hexagonal ordering is restored
after every $\tau$ time units,
where $\tau=a/(v_{\rm trap}\Delta t)$ and $\Delta t$ is the size of a simulation time step.
In Fig.~\ref{fig:1}(b) we illustrate the portion of the cycle in which the spacing between
the traps reaches its smallest value.

The sample contains
$N_{c}$ colloidal particles and we characterize the filling fraction as
$f = N_{c}/N_{\rm trap}$. 
The dynamical evolution of the colloids is given by the following overdamped
equation of motion:
\begin{equation} 
\frac{1}{\eta}\frac{\Delta {\bf r}_i}{\Delta t} =  {\bf F}_{pp}^i + {\bf F}_{\rm trap}^i 
\end{equation}
where $\eta=1$ is the viscosity.
The interaction between 
two charged colloidal particles $i$ and $j$ at a distance of ${\bf r}_{ij}$ 
is given by a screened Coulomb interaction,
${\bf F}_{pp}^{ij}=\exp{(-r/r_0)}{\bf \hat r}_{ij}/r^2$, where $r_0 =4.0$ is the screening length.
The interaction between colloid $i$ and trap $k$ is
given by 
a simple finite range harmonic spring, 
${\bf F}_{\rm trap}=(F_{\rm trap}r_{ik}/R_{\rm trap}){\bf \hat r}_{ik}$, where $F_{\rm trap}$
is the maximum force at the edge of the trap
and $r_{ik}$ is the distance between the colloid and the center of the trap.
The colloids are initialized at random locations with a specified minimum possible
spacing between adjacent colloids. 
The traps are then set into motion and the system
eventually settles into 
a steady state.

\begin{figure}
\includegraphics[width=3.5in]{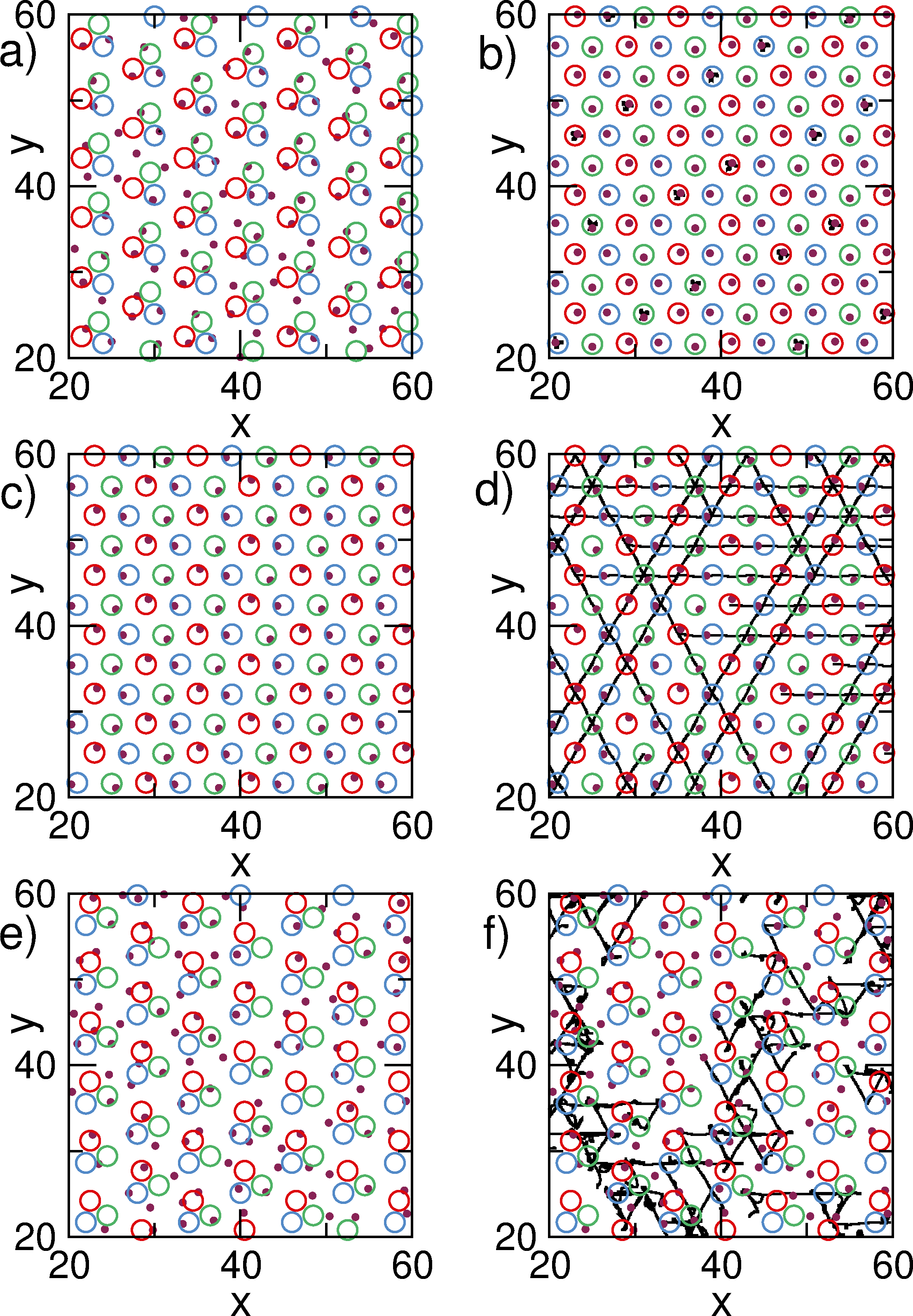}
\caption{The trap positions (open circles) and colloid positions (dots)
  in a portion of the sample at $v_{\rm trap}=0.5$ for a filling of $N_c/N_{\rm trap}=1.0$.
  (a) The weak coupling regime
  at $F_{\rm trap} = 0.4$ in the
  initial disordered state.
  (b) The same system after it has organized to a crystalline state.
  Black lines are the trajectories of a subset of the colloids
  showing that each colloid executes
 a small counterclockwise triangular loop,
  forming a dynamic chiral lattice (DCL). 
  (c) $F_{\rm trap} = 0.7$, where each trap permanently
  captures one colloid.
  (d) The same with black lines showing the trajectories of some of
  the colloids,
  which now move in straight lines and
form a choreographic lattice (ChL). 
(e)
The frustrated liquid state at $F_{\rm trap} = 0.54$,
where the system disorders and colloids can be dragged various distances as
a function of time, as indicated by the trajectories of selected colloids
shown in panel (f). 
Videos illustrating the dynamics of these phases
are available in the supplementary information \cite{M}.  
}
\label{fig:2}
\end{figure}

{\it Results--}
We first consider the weak coupling regime with $F_{\rm trap}=0.4$ and $v_{\rm trap}=0.5$
at a filling of $f=1.0$,
where individual colloids can be trapped for a short time
but move a distance less than a trap lattice constant.
In Fig.~\ref{fig:2}(a) we illustrate the colloid and trap locations at the beginning
of the simulation when the colloid positions are disordered.
After several cycles of trap motion, the colloids organize into the
crystalline state shown in Fig.~\ref{fig:2}(b), where the diffusion constant
drops to zero and the colloids move in a nonoverlapping
pattern of counterclockwise triangular loops, a state that we term a
dynamic chiral lattice (DCL).
The size of the colloidal orbits decreases with decreasing trap strength
and falls to zero when $F_{\rm trap} = 0$, where the colloids form a static hexagonal
lattice.
At $F_{\rm trap}=0.7$, shown in Fig.~\ref{fig:2}(c),
each trap permanently captures one colloid \cite{M}.
The image of the trajectories of some of the colloids in Fig.~\ref{fig:2}(d)
indicates that each colloid follows a straight line path.
Here the traps are strong enough to overcame the colloid-colloid repulsive force
even when the traps reach their point of closest approach,
so the colloids themselves form a choreographic lattice (ChL).
Although there is no 
net drift motion averaged over all of the colloids, 
individual colloids undergo ballistic motion so the diffusion constant obeys
$D \propto t$ in the ChL phase.
When $F_{\rm trap}$ is increased further, we observe the same structure and dynamics.    

At intermediate trapping strengths between the DCL and ChL states,
the system forms a partially coupled or disordered state in which each colloid
is dragged by a trap over a distance of several lattice constants before it becomes
dislodged.
We show a snapshot of this state in Fig.~\ref{fig:2}(e) for $F_{\rm trap}=0.54$,
where the colloidal positions are disordered in the steady state.
The corresponding trajectories of some of the colloids in
Fig.~\ref{fig:2}(f)
show that there is short time ballistic behavior when the colloids are dragged;
however, the longer time behavior
is diffusive with $D \propto \sqrt{t}$.

\begin{figure}
\includegraphics[width=3.5in]{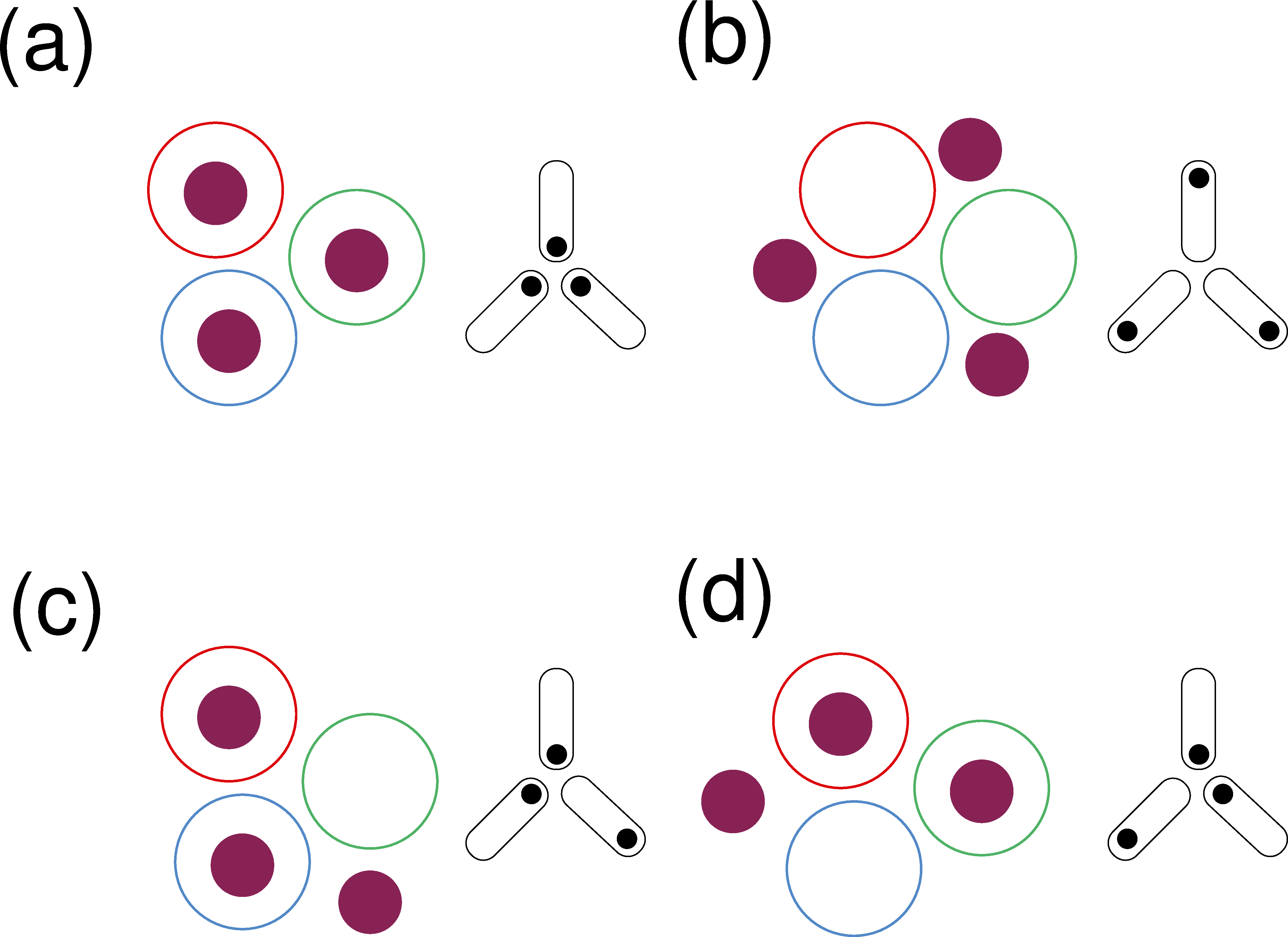}
\caption{A schematic (color) of the different vertex states in the portion of the 
choreographic crystal cycle during which the three traps are closest
together, as shown in Fig.~\ref{fig:1}(b), along with a schematic (black and white) of
the artificial spin ice vertex matching this state.
(a) When $F_{\rm trap}$ is strong enough, the colloids remain inside the 
traps, forming the ChL state in which 3-in vertices are stable.
(b) For weak $F_{\rm trap}$, all of the colloids escape from the traps,
forming a 0-in vertex state, and the system is in the DCL state.
(c,d) At intermediate $F_{\rm trap}$, 2-in vertex states are stable; however,
there are three equivalent ways to form a 2-in state, two of which are illustrated
in panels (c) and (d).  As a result, the system is frustrated and becomes disordered.
}
\label{fig:3}
\end{figure}

The mechanism that triggers the transition between the DCL and the ChL state
can be understood by considering the portion of the cycle in which the traps are
closest together, as shown in Fig.~\ref{fig:1}(b). 
We can think of this structure in terms of
a vertex picture, similar to that found in triangular colloidal spin ice systems
\cite{Libal18,Nisoli18,OrtizAmbriz19}.
A trap that is occupied is the equivalent of having a particle close to the vertex in
one of the three arms of the triangular colloidal
spin ice vertex.  Vertex states are labeled by
the total number of particles that are close to the vertex, giving 0-in, 1-in, 2-in, or 3-in
vertices.  The ice rule obeying state contains 1-in and 2-in vertices.
In Fig.~\ref{fig:3} we schematically illustrate the trap and colloid positions at the
point of closest approach, along with the corresponding
triangular colloidal spin ice vertex state.
When all three traps are occupied, as in Fig.~\ref{fig:3}(a), we have a 3-in state, which
is the highest energy configuration due to the repulsive colloid-colloid interactions.
Conversely, if all three traps are empty, as in Fig.~\ref{fig:3}(b),
the corresponding vertex is in the lowest energy 0-in state.
When the trap strength is large, the trapping energy overwhelms the colloid-colloid
interactions, stabilizing the 3-in vertex state and placing the system in the ChL phase
with ordered dynamics.
When the trap strength is weak, the colloidal interaction energy dominates
and the 0-in vertex state is stable.  Here the system is in the DCL state which
is also ordered.
For intermediate trap strength, either 1-in or 2-in vertices can form, where the 2-in
vertex has higher energy.
Whenever the trap strength is in a regime where 1-in or 2-in vertex states are favored,
the system is highly degenerate similar to the triangular colloidal spin ice.
Although there is only one possible arrangement of a system full of 3-in or 0-in vertices,
there are many possible arrangements of a system full of 1-in or 2-in vertices.  For
example, two distinct 2-in vertex configurations are illustrated in Fig.~\ref{fig:3}(c,d).
The resulting frustration at intermediate trap strength prevents
the colloids from reaching a repeatable ordered state, since each time
the traps reach their point of closest approach, a different energy-equivalent
ice-rule-obeying colloid configuration can appear.
This produces a disordered structure.
In a system with longer range 
interactions, where particles can interact over a distance spanning multiple vertices,
the ice degeneracy could be lifted, causing new types of time repeated dynamical
states to occur.
It is also possible
that other types of choreographic crystals would not have the same frustration effects
during any portion of the cycle or that
choreographic crystals could exist that are frustrated
during the entire cycle.

\begin{figure}
  \includegraphics[width=3.5in]{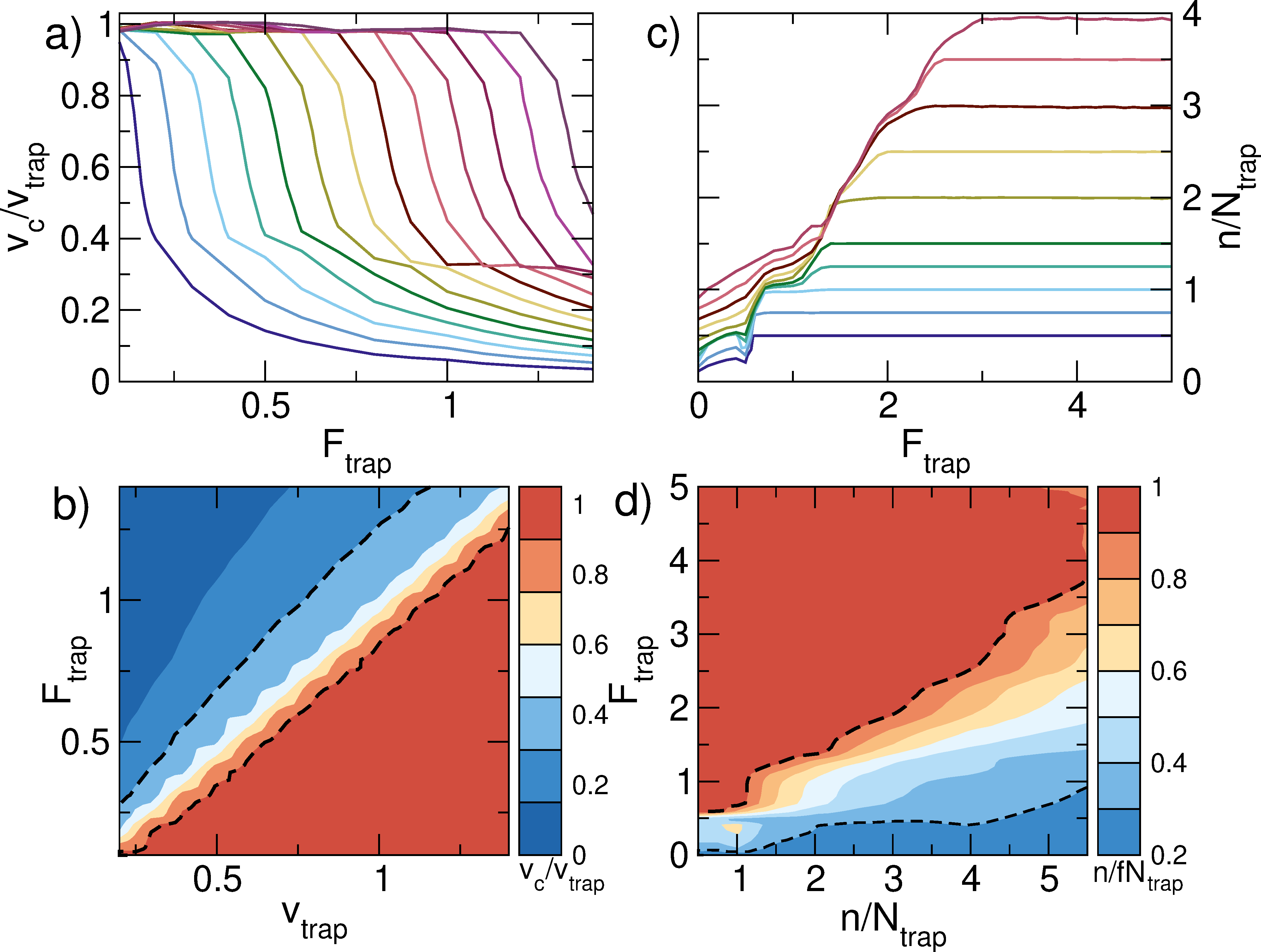}
\caption{ 
  (a) The ratio of the average colloid velocity to the trap velocity,
  $v_{c}/v_{\rm trap}$, vs $v_{\rm trap}$ for the system in Fig.~\ref{fig:2}
  with $f=1$ and $F_{\rm trap}=0.2$, 0.3, 0.4, 0.5, 0.6, 0.7, 0.8, 0.9, 1.0, 1.1, 1.2, 1.3, and 1.4,
  from bottom to top.
  The ChL phase appears  when $v_{c}/v_{\rm trap}  = 1.0$,
  the DCL phase has $v_{c}/v_{\rm trap} < 1/3$, and over the remaining
  range of $v_c/v_{\rm trap}$ the system is in the frustrated liquid state.
  As the trap speed increases, the system can pass through all three states.
  (b) The dynamic phase diagram as a function of $F_{\rm trap}$ versus $v_{\rm trap}$
  based on a height field plot of $v_c/v_{\rm trap}$
  with dashed lines marking the boundaries between
the ChL, DCL, and intermediate frustrated liquid states. 
    (c) The average number $n/N_{\rm trap}$ of colloids in each trap vs $F_{\rm trap}$
    for $v_{\rm trap}=0.5$ and
    filling fractions of $f=0.5$, 0.75, 1.0, 1.25, 1.5, 2.0, 2.5, 3.0, 3.5, and 4.0,
    from bottom to top.
    (d) The dynamic phase diagram as a function of $F_{\rm trap}$ versus $n/N_{\rm trap}$
    based on a height field plot of $n/(fN_{\rm trap})$.
}
\label{fig:4}
\end{figure}

Transitions between the states can occur as a function of trap speed as well as
trap strength, since the coupling between the colloids and the traps becomes weaker
as the trap speed increases.
In Fig.~\ref{fig:4}(a) we plot the ratio of the velocities of the
colloids $v_{c}/v_{\rm trap}=N_{c}^{-1}\sum_{i}^{N_c}|{\bf v}_i|/v_{\rm trap}$
versus $v_{\rm trap}$ for the
system in Fig.~\ref{fig:2}.
When $v_{c}/v_{\rm trap} = 1.0$, all the colloids are trapped
and moving at the same velocity as the traps in the ChL state.
For $v_{c}/v_{\rm trap} < 0.3$, the colloids are only temporarily trapped and
the DCL state appears,
while for $0.3 < v_{c}/v_{\rm trap} < 0.9$, the system is 
in the disordered state.
We find a plateau
near $v_{c}/v_{\rm trap} = 0.33$ corresponding to a
prevalence of 1-in states with 1/3 of the traps occupied.
Even in the DCL state, $v_c/v_{\rm trap}$ is always larger than zero since the traps
are occupied for at least a short period of time.
As $F_{\rm trap}$ increases, the onset of the ChL phase shifts to lower values
of $v_{\rm trap}$.
In Fig.~\ref{fig:4}(b) we plot a dynamic phase diagram as a function of
$F_{\rm trap}$ vs $v_{\rm trap}$
for the system in Fig.~\ref{fig:4}(a) indicating the regions where the
DCL, ChL, and  disordered phases occur.

We have also explored the effect of changing the filling $f$.
We find the same general features described above when $f\leq 1$,
while interstitial colloids begin to appear when $f>1$.
If the trap strength is large enough, however, all the colloids can eventually be trapped
and the ChL phase appears with
clusters of colloids at each trap.
The time-averaged number of trapped colloids is given by $n$, so the
average number of colloids in each trap is $n/N_{\rm trap}$.
When the traps are not strong enough to capture an average of
$n/N_{\rm trap}=f$ colloids apiece,
more complex states can appear in which multiply occupied traps coexist with
interstitial colloids.
In general, the region over which the disordered phase
appears expands as $f$ increases.
In Fig.~\ref{fig:4}(c) we plot $n/N_{\rm trap}$ versus 
$F_{\rm trap}$ for varied filling.
A series of plateaus occur at $n/N_{\rm trap}=1/f$ when the traps are strong enough
to capture all the colloids with no interstitial colloids present.
In Fig.~\ref{fig:4}(d) we plot a dynamic phase diagram as a function of
$F_{\rm trap}$ versus $n/N_{\rm trap}$
highlighting the strongly coupled regime or ChL phase,
the weakly coupled DCL regime, and the intermediate regime consisting of
disordered states in which multiply trapped colloids can coexist with
interstitial colloids.
 
{\it Summary---}
We have examined a choreographic lattice of traps that move in a synchronized fashion
without overlap.
When the traps are coupled 
to an assembly of colloidal particles with repulsive Yukawa interactions,
we observe several different dynamical regimes:
a dynamically ordered chiral crystal state in which the colloids are temporarily trapped,
follow loop orbits, and have zero net diffusion; a strongly coupled state in which
the colloids themselves form a choreographic lattice
with ballistic diffusion; and an intermediate frustrated liquid state
with long time diffusive behavior.
The emergence of the different states can be understood in terms of
a mapping of the closest approach of the traps
to a triangular colloidal spin ice vertex state.
At intermediate coupling, multiple vertex states with equivalent energies
are possible, resulting in frustration similar to that found in the triangular
colloidal spin ice, which produces a disordered configuration.
Our results could be generalized to a wide variety of different types
of choreographic time crystals with dynamical substrates,
and represent a new particle assembly-periodic substrate system
in which commensuration effects, dynamic phases,
and melting can be explored using optical traps or other methods to
create a translating trap array. Beyond colloids,
similar results should appear for vortices, cold atoms, and ions coupled to a choreographic lattice.    

\acknowledgments
This work was supported by the US Department of Energy through
the Los Alamos National Laboratory.  Los Alamos National Laboratory is
operated by Triad National Security, LLC, for the National Nuclear Security
Administration of the U. S. Department of Energy (Contract No. 892333218NCA000001).

\bibliography{mybib}

\begin{thebibliography}{44}%
\makeatletter
\providecommand \@ifxundefined [1]{%
 \@ifx{#1\undefined}
}%
\providecommand \@ifnum [1]{%
 \ifnum #1\expandafter \@firstoftwo
 \else \expandafter \@secondoftwo
 \fi
}%
\providecommand \@ifx [1]{%
 \ifx #1\expandafter \@firstoftwo
 \else \expandafter \@secondoftwo
 \fi
}%
\providecommand \natexlab [1]{#1}%
\providecommand \enquote  [1]{``#1''}%
\providecommand \bibnamefont  [1]{#1}%
\providecommand \bibfnamefont [1]{#1}%
\providecommand \citenamefont [1]{#1}%
\providecommand \href@noop [0]{\@secondoftwo}%
\providecommand \href [0]{\begingroup \@sanitize@url \@href}%
\providecommand \@href[1]{\@@startlink{#1}\@@href}%
\providecommand \@@href[1]{\endgroup#1\@@endlink}%
\providecommand \@sanitize@url [0]{\catcode `\\12\catcode `\$12\catcode
  `\&12\catcode `\#12\catcode `\^12\catcode `\_12\catcode `\%12\relax}%
\providecommand \@@startlink[1]{}%
\providecommand \@@endlink[0]{}%
\providecommand \url  [0]{\begingroup\@sanitize@url \@url }%
\providecommand \@url [1]{\endgroup\@href {#1}{\urlprefix }}%
\providecommand \urlprefix  [0]{URL }%
\providecommand \Eprint [0]{\href }%
\providecommand \doibase [0]{http://dx.doi.org/}%
\providecommand \selectlanguage [0]{\@gobble}%
\providecommand \bibinfo  [0]{\@secondoftwo}%
\providecommand \bibfield  [0]{\@secondoftwo}%
\providecommand \translation [1]{[#1]}%
\providecommand \BibitemOpen [0]{}%
\providecommand \bibitemStop [0]{}%
\providecommand \bibitemNoStop [0]{.\EOS\space}%
\providecommand \EOS [0]{\spacefactor3000\relax}%
\providecommand \BibitemShut  [1]{\csname bibitem#1\endcsname}%
\let\auto@bib@innerbib\@empty
%</preamble>
\bibitem [{\citenamefont {Boyle}\ \emph {et~al.}(2016)\citenamefont {Boyle},
  \citenamefont {Khoo},\ and\ \citenamefont {Smith}}]{Boyle16}%
  \BibitemOpen
  \bibfield  {author} {\bibinfo {author} {\bibfnamefont {L.}~\bibnamefont
  {Boyle}}, \bibinfo {author} {\bibfnamefont {J.~Y.}\ \bibnamefont {Khoo}}, \
  and\ \bibinfo {author} {\bibfnamefont {K.}~\bibnamefont {Smith}},\ }\bibfield
   {title} {\enquote {\bibinfo {title} {Symmetric satellite swarms and
  choreographic crystals},}\ }\href {\doibase 10.1103/PhysRevLett.116.015503}
  {\bibfield  {journal} {\bibinfo  {journal} {Phys. Rev. Lett.}\ }\textbf
  {\bibinfo {volume} {116}},\ \bibinfo {pages} {015503} (\bibinfo {year}
  {2016})}\BibitemShut {NoStop}%
\bibitem [{\citenamefont {Shapere}\ and\ \citenamefont
  {Wilczek}(2012)}]{Shapere12}%
  \BibitemOpen
  \bibfield  {author} {\bibinfo {author} {\bibfnamefont {A.}~\bibnamefont
  {Shapere}}\ and\ \bibinfo {author} {\bibfnamefont {F.}~\bibnamefont
  {Wilczek}},\ }\bibfield  {title} {\enquote {\bibinfo {title} {Classical time
  crystals},}\ }\href {\doibase 10.1103/PhysRevLett.109.160402} {\bibfield
  {journal} {\bibinfo  {journal} {Phys. Rev. Lett.}\ }\textbf {\bibinfo
  {volume} {109}},\ \bibinfo {pages} {160402} (\bibinfo {year}
  {2012})}\BibitemShut {NoStop}%
\bibitem [{\citenamefont {Sacha}\ and\ \citenamefont
  {Zakrzewski}(2018)}]{Sacha18}%
  \BibitemOpen
  \bibfield  {author} {\bibinfo {author} {\bibfnamefont {K.}~\bibnamefont
  {Sacha}}\ and\ \bibinfo {author} {\bibfnamefont {J.}~\bibnamefont
  {Zakrzewski}},\ }\bibfield  {title} {\enquote {\bibinfo {title} {Time
  crystals: a review},}\ }\href {\doibase 10.1088/1361-6633/aa8b38} {\bibfield
  {journal} {\bibinfo  {journal} {Rep. Prog. Phys.}\ }\textbf {\bibinfo
  {volume} {81}},\ \bibinfo {pages} {016401} (\bibinfo {year}
  {2018})}\BibitemShut {NoStop}%
\bibitem [{\citenamefont {Flicker}(2018)}]{Flicker18}%
  \BibitemOpen
  \bibfield  {author} {\bibinfo {author} {\bibfnamefont {F.}~\bibnamefont
  {Flicker}},\ }\bibfield  {title} {\enquote {\bibinfo {title} {Time
  quasilattices in dissipative dynamical systems},}\ }\href {\doibase
  10.21468/SciPostPhys.5.1.001} {\bibfield  {journal} {\bibinfo  {journal}
  {SciPost Phys.}\ }\textbf {\bibinfo {volume} {5}},\ \bibinfo {pages} {001}
  (\bibinfo {year} {2018})}\BibitemShut {NoStop}%
\bibitem [{\citenamefont {Yao}\ \emph {et~al.}()\citenamefont {Yao},
  \citenamefont {Nayak}, \citenamefont {Balents},\ and\ \citenamefont
  {Zaletel}}]{Yao18}%
  \BibitemOpen
  \bibfield  {author} {\bibinfo {author} {\bibfnamefont {N.~Y.}\ \bibnamefont
  {Yao}}, \bibinfo {author} {\bibfnamefont {C.}~\bibnamefont {Nayak}}, \bibinfo
  {author} {\bibfnamefont {L.}~\bibnamefont {Balents}}, \ and\ \bibinfo
  {author} {\bibfnamefont {M.~P.}\ \bibnamefont {Zaletel}},\ }\href@noop {}
  {\enquote {\bibinfo {title} {Classical discrete time crystals},}\ }\bibinfo
  {note} {{arXiv:1801.02628}}\BibitemShut {NoStop}%
\bibitem [{\citenamefont {Dai}\ \emph {et~al.}(2019)\citenamefont {Dai},
  \citenamefont {Niemi}, \citenamefont {Peng},\ and\ \citenamefont
  {Wilczek}}]{Dai19}%
  \BibitemOpen
  \bibfield  {author} {\bibinfo {author} {\bibfnamefont {J.}~\bibnamefont
  {Dai}}, \bibinfo {author} {\bibfnamefont {A.~J.}\ \bibnamefont {Niemi}},
  \bibinfo {author} {\bibfnamefont {X.}~\bibnamefont {Peng}}, \ and\ \bibinfo
  {author} {\bibfnamefont {F.}~\bibnamefont {Wilczek}},\ }\bibfield  {title}
  {\enquote {\bibinfo {title} {Truncated dynamics, ring molecules, and
  mechanical time crystals},}\ }\href {\doibase 10.1103/PhysRevA.99.023425}
  {\bibfield  {journal} {\bibinfo  {journal} {Phys. Rev. A}\ }\textbf {\bibinfo
  {volume} {99}},\ \bibinfo {pages} {023425} (\bibinfo {year}
  {2019})}\BibitemShut {NoStop}%
\bibitem [{\citenamefont {Wilczek}(2012)}]{Wilczek12}%
  \BibitemOpen
  \bibfield  {author} {\bibinfo {author} {\bibfnamefont {F.}~\bibnamefont
  {Wilczek}},\ }\bibfield  {title} {\enquote {\bibinfo {title} {Quantum time
  crystals},}\ }\href {\doibase 10.1103/PhysRevLett.109.160401} {\bibfield
  {journal} {\bibinfo  {journal} {Phys. Rev. Lett.}\ }\textbf {\bibinfo
  {volume} {109}},\ \bibinfo {pages} {160401} (\bibinfo {year}
  {2012})}\BibitemShut {NoStop}%
\bibitem [{\citenamefont {Li}\ \emph {et~al.}(2012)\citenamefont {Li},
  \citenamefont {Gong}, \citenamefont {Yin}, \citenamefont {Quan},
  \citenamefont {Yin}, \citenamefont {Zhang}, \citenamefont {Duan},\ and\
  \citenamefont {Zhang}}]{Li12}%
  \BibitemOpen
  \bibfield  {author} {\bibinfo {author} {\bibfnamefont {T.}~\bibnamefont
  {Li}}, \bibinfo {author} {\bibfnamefont {Z.-X.}\ \bibnamefont {Gong}},
  \bibinfo {author} {\bibfnamefont {Z.-Q.}\ \bibnamefont {Yin}}, \bibinfo
  {author} {\bibfnamefont {H.~T.}\ \bibnamefont {Quan}}, \bibinfo {author}
  {\bibfnamefont {X.}~\bibnamefont {Yin}}, \bibinfo {author} {\bibfnamefont
  {P.}~\bibnamefont {Zhang}}, \bibinfo {author} {\bibfnamefont {L.-M.}\
  \bibnamefont {Duan}}, \ and\ \bibinfo {author} {\bibfnamefont
  {X.}~\bibnamefont {Zhang}},\ }\bibfield  {title} {\enquote {\bibinfo {title}
  {Space-time crystals of trapped ions},}\ }\href {\doibase
  10.1103/PhysRevLett.109.163001} {\bibfield  {journal} {\bibinfo  {journal}
  {Phys. Rev. Lett.}\ }\textbf {\bibinfo {volume} {109}},\ \bibinfo {pages}
  {163001} (\bibinfo {year} {2012})}\BibitemShut {NoStop}%
\bibitem [{\citenamefont {Zhang}\ \emph {et~al.}(2017)\citenamefont {Zhang},
  \citenamefont {Hess}, \citenamefont {Kyprianidis}, \citenamefont {Becker},
  \citenamefont {Lee}, \citenamefont {Smith}, \citenamefont {Pagano},
  \citenamefont {Potirniche}, \citenamefont {Potter}, \citenamefont
  {Vishwanath}, \citenamefont {Yao},\ and\ \citenamefont {Monroe}}]{Zhang17}%
  \BibitemOpen
  \bibfield  {author} {\bibinfo {author} {\bibfnamefont {J.}~\bibnamefont
  {Zhang}}, \bibinfo {author} {\bibfnamefont {P.~W.}\ \bibnamefont {Hess}},
  \bibinfo {author} {\bibfnamefont {A.}~\bibnamefont {Kyprianidis}}, \bibinfo
  {author} {\bibfnamefont {P.}~\bibnamefont {Becker}}, \bibinfo {author}
  {\bibfnamefont {A.}~\bibnamefont {Lee}}, \bibinfo {author} {\bibfnamefont
  {J.}~\bibnamefont {Smith}}, \bibinfo {author} {\bibfnamefont
  {G.}~\bibnamefont {Pagano}}, \bibinfo {author} {\bibfnamefont {I.~D.}\
  \bibnamefont {Potirniche}}, \bibinfo {author} {\bibfnamefont {A.~C.}\
  \bibnamefont {Potter}}, \bibinfo {author} {\bibfnamefont {A.}~\bibnamefont
  {Vishwanath}}, \bibinfo {author} {\bibfnamefont {N.~Y.}\ \bibnamefont {Yao}},
  \ and\ \bibinfo {author} {\bibfnamefont {C.}~\bibnamefont {Monroe}},\
  }\bibfield  {title} {\enquote {\bibinfo {title} {Observation of a discrete
  time crystal},}\ }\href {\doibase 10.1038/nature21413} {\bibfield  {journal}
  {\bibinfo  {journal} {Nature (London)}\ }\textbf {\bibinfo {volume} {543}},\
  \bibinfo {pages} {217} (\bibinfo {year} {2017})}\BibitemShut {NoStop}%
\bibitem [{\citenamefont {Bo\ifmmode \check{s}\else
  \v{s}\fi{}kovi\ifmmode~\acute{c}\else \'{c}\fi{}}\ \emph
  {et~al.}(2018)\citenamefont {Bo\ifmmode \check{s}\else
  \v{s}\fi{}kovi\ifmmode~\acute{c}\else \'{c}\fi{}}, \citenamefont {Duque},
  \citenamefont {Ferreira}, \citenamefont {Miguel},\ and\ \citenamefont
  {Cardoso}}]{Boskovic18}%
  \BibitemOpen
  \bibfield  {author} {\bibinfo {author} {\bibfnamefont {M.}~\bibnamefont
  {Bo\ifmmode \check{s}\else \v{s}\fi{}kovi\ifmmode~\acute{c}\else
  \'{c}\fi{}}}, \bibinfo {author} {\bibfnamefont {F.}~\bibnamefont {Duque}},
  \bibinfo {author} {\bibfnamefont {M.~C.}\ \bibnamefont {Ferreira}}, \bibinfo
  {author} {\bibfnamefont {F.~S.}\ \bibnamefont {Miguel}}, \ and\ \bibinfo
  {author} {\bibfnamefont {V.}~\bibnamefont {Cardoso}},\ }\bibfield  {title}
  {\enquote {\bibinfo {title} {Motion in time-periodic backgrounds with
  applications to ultralight dark matter halos at galactic centers},}\ }\href
  {\doibase 10.1103/PhysRevD.98.024037} {\bibfield  {journal} {\bibinfo
  {journal} {Phys. Rev. D}\ }\textbf {\bibinfo {volume} {98}},\ \bibinfo
  {pages} {024037} (\bibinfo {year} {2018})}\BibitemShut {NoStop}%
\bibitem [{\citenamefont {Easson}\ and\ \citenamefont
  {Manton}(2019)}]{Easson19}%
  \BibitemOpen
  \bibfield  {author} {\bibinfo {author} {\bibfnamefont {D.~A.}\ \bibnamefont
  {Easson}}\ and\ \bibinfo {author} {\bibfnamefont {T.}~\bibnamefont
  {Manton}},\ }\bibfield  {title} {\enquote {\bibinfo {title} {Stable cosmic
  time crystals},}\ }\href {\doibase 10.1103/PhysRevD.99.043507} {\bibfield
  {journal} {\bibinfo  {journal} {Phys. Rev. D}\ }\textbf {\bibinfo {volume}
  {99}},\ \bibinfo {pages} {043507} (\bibinfo {year} {2019})}\BibitemShut
  {NoStop}%
\bibitem [{\citenamefont {Chowdhury}\ \emph {et~al.}(1985)\citenamefont
  {Chowdhury}, \citenamefont {Ackerson},\ and\ \citenamefont
  {Clark}}]{Chowdhury85}%
  \BibitemOpen
  \bibfield  {author} {\bibinfo {author} {\bibfnamefont {A.}~\bibnamefont
  {Chowdhury}}, \bibinfo {author} {\bibfnamefont {B.~J.}\ \bibnamefont
  {Ackerson}}, \ and\ \bibinfo {author} {\bibfnamefont {N.~A.}\ \bibnamefont
  {Clark}},\ }\bibfield  {title} {\enquote {\bibinfo {title} {Laser-induced
  freezing},}\ }\href {\doibase 10.1103/PhysRevLett.55.833} {\bibfield
  {journal} {\bibinfo  {journal} {Phys. Rev. Lett.}\ }\textbf {\bibinfo
  {volume} {55}},\ \bibinfo {pages} {833--836} (\bibinfo {year}
  {1985})}\BibitemShut {NoStop}%
\bibitem [{\citenamefont {Wei}\ \emph {et~al.}(1998)\citenamefont {Wei},
  \citenamefont {Bechinger}, \citenamefont {Rudhardt},\ and\ \citenamefont
  {Leiderer}}]{Wei98}%
  \BibitemOpen
  \bibfield  {author} {\bibinfo {author} {\bibfnamefont {Q.-H.}\ \bibnamefont
  {Wei}}, \bibinfo {author} {\bibfnamefont {C.}~\bibnamefont {Bechinger}},
  \bibinfo {author} {\bibfnamefont {D.}~\bibnamefont {Rudhardt}}, \ and\
  \bibinfo {author} {\bibfnamefont {P.}~\bibnamefont {Leiderer}},\ }\bibfield
  {title} {\enquote {\bibinfo {title} {Experimental study of laser-induced
  melting in two-dimensional colloids},}\ }\href {\doibase
  10.1103/PhysRevLett.81.2606} {\bibfield  {journal} {\bibinfo  {journal}
  {Phys. Rev. Lett.}\ }\textbf {\bibinfo {volume} {81}},\ \bibinfo {pages}
  {2606--2609} (\bibinfo {year} {1998})}\BibitemShut {NoStop}%
\bibitem [{\citenamefont {Korda}\ \emph
  {et~al.}(2002{\natexlab{a}})\citenamefont {Korda}, \citenamefont {Spalding},\
  and\ \citenamefont {Grier}}]{Korda02a}%
  \BibitemOpen
  \bibfield  {author} {\bibinfo {author} {\bibfnamefont {P.~T.}\ \bibnamefont
  {Korda}}, \bibinfo {author} {\bibfnamefont {G.~C.}\ \bibnamefont {Spalding}},
  \ and\ \bibinfo {author} {\bibfnamefont {D.~G.}\ \bibnamefont {Grier}},\
  }\bibfield  {title} {\enquote {\bibinfo {title} {Evolution of a colloidal
  critical state in an optical pinning potential landscape},}\ }\href {\doibase
  10.1103/PhysRevB.66.024504} {\bibfield  {journal} {\bibinfo  {journal} {Phys.
  Rev. B}\ }\textbf {\bibinfo {volume} {66}},\ \bibinfo {pages} {024504}
  (\bibinfo {year} {2002}{\natexlab{a}})}\BibitemShut {NoStop}%
\bibitem [{\citenamefont {Reichhardt}\ and\ \citenamefont
  {Olson}(2002)}]{Reichhardt02a}%
  \BibitemOpen
  \bibfield  {author} {\bibinfo {author} {\bibfnamefont {C.}~\bibnamefont
  {Reichhardt}}\ and\ \bibinfo {author} {\bibfnamefont {C.~J.}\ \bibnamefont
  {Olson}},\ }\bibfield  {title} {\enquote {\bibinfo {title} {Novel colloidal
  crystalline states on two-dimensional periodic substrates},}\ }\href
  {\doibase 10.1103/PhysRevLett.88.248301} {\bibfield  {journal} {\bibinfo
  {journal} {Phys. Rev. Lett.}\ }\textbf {\bibinfo {volume} {88}},\ \bibinfo
  {pages} {248301} (\bibinfo {year} {2002})}\BibitemShut {NoStop}%
\bibitem [{\citenamefont {Brunner}\ and\ \citenamefont
  {Bechinger}(2002)}]{Brunner02}%
  \BibitemOpen
  \bibfield  {author} {\bibinfo {author} {\bibfnamefont {M.}~\bibnamefont
  {Brunner}}\ and\ \bibinfo {author} {\bibfnamefont {C.}~\bibnamefont
  {Bechinger}},\ }\bibfield  {title} {\enquote {\bibinfo {title} {Phase
  behavior of colloidal molecular crystals on triangular light lattices},}\
  }\href {\doibase 10.1103/PhysRevLett.88.248302} {\bibfield  {journal}
  {\bibinfo  {journal} {Phys. Rev. Lett.}\ }\textbf {\bibinfo {volume} {88}},\
  \bibinfo {pages} {248302} (\bibinfo {year} {2002})}\BibitemShut {NoStop}%
\bibitem [{\citenamefont {Agra}\ \emph {et~al.}(2004)\citenamefont {Agra},
  \citenamefont {van Wijland},\ and\ \citenamefont {Trizac}}]{Agra04}%
  \BibitemOpen
  \bibfield  {author} {\bibinfo {author} {\bibfnamefont {R.}~\bibnamefont
  {Agra}}, \bibinfo {author} {\bibfnamefont {F.}~\bibnamefont {van Wijland}}, \
  and\ \bibinfo {author} {\bibfnamefont {E.}~\bibnamefont {Trizac}},\
  }\bibfield  {title} {\enquote {\bibinfo {title} {Theory of orientational
  ordering in colloidal molecular crystals},}\ }\href {\doibase
  10.1103/PhysRevLett.93.018304} {\bibfield  {journal} {\bibinfo  {journal}
  {Phys. Rev. Lett.}\ }\textbf {\bibinfo {volume} {93}},\ \bibinfo {pages}
  {018304} (\bibinfo {year} {2004})}\BibitemShut {NoStop}%
\bibitem [{\citenamefont {Ortiz-Ambriz}\ and\ \citenamefont
  {Tierno}(2016)}]{OrtizAmbriz16}%
  \BibitemOpen
  \bibfield  {author} {\bibinfo {author} {\bibfnamefont {A.}~\bibnamefont
  {Ortiz-Ambriz}}\ and\ \bibinfo {author} {\bibfnamefont {P.}~\bibnamefont
  {Tierno}},\ }\bibfield  {title} {\enquote {\bibinfo {title} {Engineering of
  frustration in colloidal artificial ices realized on microfeatured grooved
  lattices},}\ }\href {\doibase 10.1038/ncomms10575} {\bibfield  {journal}
  {\bibinfo  {journal} {Nature Commun.}\ }\textbf {\bibinfo {volume} {7}},\
  \bibinfo {pages} {10575} (\bibinfo {year} {2016})}\BibitemShut {NoStop}%
\bibitem [{\citenamefont {Brazda}\ \emph {et~al.}(2018)\citenamefont {Brazda},
  \citenamefont {Silva}, \citenamefont {Manini}, \citenamefont {Vanossi},
  \citenamefont {Guerra}, \citenamefont {Tosatti},\ and\ \citenamefont
  {Bechinger}}]{Brazda18}%
  \BibitemOpen
  \bibfield  {author} {\bibinfo {author} {\bibfnamefont {T.}~\bibnamefont
  {Brazda}}, \bibinfo {author} {\bibfnamefont {A.}~\bibnamefont {Silva}},
  \bibinfo {author} {\bibfnamefont {N.}~\bibnamefont {Manini}}, \bibinfo
  {author} {\bibfnamefont {A.}~\bibnamefont {Vanossi}}, \bibinfo {author}
  {\bibfnamefont {R.}~\bibnamefont {Guerra}}, \bibinfo {author} {\bibfnamefont
  {E.}~\bibnamefont {Tosatti}}, \ and\ \bibinfo {author} {\bibfnamefont
  {C.}~\bibnamefont {Bechinger}},\ }\bibfield  {title} {\enquote {\bibinfo
  {title} {Experimental observation of the {A}ubry transition in
  two-dimensional colloidal monolayers},}\ }\href {\doibase
  10.1103/PhysRevX.8.011050} {\bibfield  {journal} {\bibinfo  {journal} {Phys.
  Rev. X}\ }\textbf {\bibinfo {volume} {8}},\ \bibinfo {pages} {011050}
  (\bibinfo {year} {2018})}\BibitemShut {NoStop}%
\bibitem [{\citenamefont {Mikhael}\ \emph {et~al.}(2008)\citenamefont
  {Mikhael}, \citenamefont {Roth}, \citenamefont {Helden},\ and\ \citenamefont
  {Bechinger}}]{Mikhael08}%
  \BibitemOpen
  \bibfield  {author} {\bibinfo {author} {\bibfnamefont {J.}~\bibnamefont
  {Mikhael}}, \bibinfo {author} {\bibfnamefont {J.}~\bibnamefont {Roth}},
  \bibinfo {author} {\bibfnamefont {L.}~\bibnamefont {Helden}}, \ and\ \bibinfo
  {author} {\bibfnamefont {C.}~\bibnamefont {Bechinger}},\ }\bibfield  {title}
  {\enquote {\bibinfo {title} {Archimedean-like tiling on decagonal
  quasicrystalline surfaces},}\ }\href {\doibase 10.1038/nature07074}
  {\bibfield  {journal} {\bibinfo  {journal} {Nature (London)}\ }\textbf
  {\bibinfo {volume} {454}},\ \bibinfo {pages} {501--504} (\bibinfo {year}
  {2008})}\BibitemShut {NoStop}%
\bibitem [{\citenamefont {Schmiedeberg}\ and\ \citenamefont
  {Stark}(2008)}]{Schmiedeberg08}%
  \BibitemOpen
  \bibfield  {author} {\bibinfo {author} {\bibfnamefont {M.}~\bibnamefont
  {Schmiedeberg}}\ and\ \bibinfo {author} {\bibfnamefont {H.}~\bibnamefont
  {Stark}},\ }\bibfield  {title} {\enquote {\bibinfo {title} {Colloidal
  ordering on a {2D} quasicrystalline substrate},}\ }\href {\doibase
  10.1103/PhysRevLett.101.218302} {\bibfield  {journal} {\bibinfo  {journal}
  {Phys. Rev. Lett.}\ }\textbf {\bibinfo {volume} {101}},\ \bibinfo {pages}
  {218302} (\bibinfo {year} {2008})}\BibitemShut {NoStop}%
\bibitem [{\citenamefont {Korda}\ \emph
  {et~al.}(2002{\natexlab{b}})\citenamefont {Korda}, \citenamefont {Taylor},\
  and\ \citenamefont {Grier}}]{Korda02}%
  \BibitemOpen
  \bibfield  {author} {\bibinfo {author} {\bibfnamefont {P.~T.}\ \bibnamefont
  {Korda}}, \bibinfo {author} {\bibfnamefont {M.~B.}\ \bibnamefont {Taylor}}, \
  and\ \bibinfo {author} {\bibfnamefont {D.~G.}\ \bibnamefont {Grier}},\
  }\bibfield  {title} {\enquote {\bibinfo {title} {Kinetically locked-in
  colloidal transport in an array of optical tweezers},}\ }\href {\doibase
  10.1103/PhysRevLett.89.128301} {\bibfield  {journal} {\bibinfo  {journal}
  {Phys. Rev. Lett.}\ }\textbf {\bibinfo {volume} {89}},\ \bibinfo {pages}
  {128301} (\bibinfo {year} {2002}{\natexlab{b}})}\BibitemShut {NoStop}%
\bibitem [{\citenamefont {MacDonald}\ \emph {et~al.}(2003)\citenamefont
  {MacDonald}, \citenamefont {Spalding},\ and\ \citenamefont
  {Dholakia}}]{MacDonald03}%
  \BibitemOpen
  \bibfield  {author} {\bibinfo {author} {\bibfnamefont {M.~P.}\ \bibnamefont
  {MacDonald}}, \bibinfo {author} {\bibfnamefont {G.~C.}\ \bibnamefont
  {Spalding}}, \ and\ \bibinfo {author} {\bibfnamefont {K.}~\bibnamefont
  {Dholakia}},\ }\bibfield  {title} {\enquote {\bibinfo {title} {Microfluidic
  sorting in an optical lattice},}\ }\href {\doibase 10.1038/nature02144}
  {\bibfield  {journal} {\bibinfo  {journal} {Nature (London)}\ }\textbf
  {\bibinfo {volume} {426}},\ \bibinfo {pages} {421--424} (\bibinfo {year}
  {2003})}\BibitemShut {NoStop}%
\bibitem [{\citenamefont {Reichhardt}\ and\ \citenamefont
  {Olson~Reichhardt}(2011)}]{Reichhardt11}%
  \BibitemOpen
  \bibfield  {author} {\bibinfo {author} {\bibfnamefont {C.}~\bibnamefont
  {Reichhardt}}\ and\ \bibinfo {author} {\bibfnamefont {C.~J.}\ \bibnamefont
  {Olson~Reichhardt}},\ }\bibfield  {title} {\enquote {\bibinfo {title}
  {Dynamical ordering and directional locking for particles moving over
  quasicrystalline substrates},}\ }\href {\doibase
  10.1103/PhysRevLett.106.060603} {\bibfield  {journal} {\bibinfo  {journal}
  {Phys. Rev. Lett.}\ }\textbf {\bibinfo {volume} {106}},\ \bibinfo {pages}
  {060603} (\bibinfo {year} {2011})}\BibitemShut {NoStop}%
\bibitem [{\citenamefont {Bohlein}\ and\ \citenamefont
  {Bechinger}(2012)}]{Bohlein12a}%
  \BibitemOpen
  \bibfield  {author} {\bibinfo {author} {\bibfnamefont {T.}~\bibnamefont
  {Bohlein}}\ and\ \bibinfo {author} {\bibfnamefont {C.}~\bibnamefont
  {Bechinger}},\ }\bibfield  {title} {\enquote {\bibinfo {title} {Experimental
  observation of directional locking and dynamical ordering of colloidal
  monolayers driven across quasiperiodic substrates},}\ }\href {\doibase
  10.1103/PhysRevLett.109.058301} {\bibfield  {journal} {\bibinfo  {journal}
  {Phys. Rev. Lett.}\ }\textbf {\bibinfo {volume} {109}},\ \bibinfo {pages}
  {058301} (\bibinfo {year} {2012})}\BibitemShut {NoStop}%
\bibitem [{\citenamefont {Cao}\ \emph {et~al.}(2019)\citenamefont {Cao},
  \citenamefont {Panizon}, \citenamefont {Vanossi}, \citenamefont {Manini},\
  and\ \citenamefont {Bechinger}}]{Cao19}%
  \BibitemOpen
  \bibfield  {author} {\bibinfo {author} {\bibfnamefont {X.}~\bibnamefont
  {Cao}}, \bibinfo {author} {\bibfnamefont {E.}~\bibnamefont {Panizon}},
  \bibinfo {author} {\bibfnamefont {A.}~\bibnamefont {Vanossi}}, \bibinfo
  {author} {\bibfnamefont {N.}~\bibnamefont {Manini}}, \ and\ \bibinfo {author}
  {\bibfnamefont {C.}~\bibnamefont {Bechinger}},\ }\bibfield  {title} {\enquote
  {\bibinfo {title} {Orientational and directional locking of colloidal
  clusters driven across periodic surfaces},}\ }\href {\doibase
  10.1038/s41567-019-0515-7} {\bibfield  {journal} {\bibinfo  {journal} {Nature
  Phys.}\ }\textbf {\bibinfo {volume} {15}},\ \bibinfo {pages} {776} (\bibinfo
  {year} {2019})}\BibitemShut {NoStop}%
\bibitem [{\citenamefont {Bohlein}\ \emph {et~al.}(2012)\citenamefont
  {Bohlein}, \citenamefont {Mikhael},\ and\ \citenamefont
  {Bechinger}}]{Bohlein12}%
  \BibitemOpen
  \bibfield  {author} {\bibinfo {author} {\bibfnamefont {T.}~\bibnamefont
  {Bohlein}}, \bibinfo {author} {\bibfnamefont {J.}~\bibnamefont {Mikhael}}, \
  and\ \bibinfo {author} {\bibfnamefont {C.}~\bibnamefont {Bechinger}},\
  }\bibfield  {title} {\enquote {\bibinfo {title} {Observation of kinks and
  antikinks in colloidal monolayers driven across ordered surfaces},}\ }\href
  {\doibase 10.1038/NMAT3204} {\bibfield  {journal} {\bibinfo  {journal}
  {Nature Mater.}\ }\textbf {\bibinfo {volume} {11}},\ \bibinfo {pages}
  {126--130} (\bibinfo {year} {2012})}\BibitemShut {NoStop}%
\bibitem [{\citenamefont {Vanossi}\ \emph {et~al.}(2012)\citenamefont
  {Vanossi}, \citenamefont {Manini},\ and\ \citenamefont
  {Tosatti}}]{Vanossi12}%
  \BibitemOpen
  \bibfield  {author} {\bibinfo {author} {\bibfnamefont {A.}~\bibnamefont
  {Vanossi}}, \bibinfo {author} {\bibfnamefont {N.}~\bibnamefont {Manini}}, \
  and\ \bibinfo {author} {\bibfnamefont {E.}~\bibnamefont {Tosatti}},\
  }\bibfield  {title} {\enquote {\bibinfo {title} {Static and dynamic friction
  in sliding colloidal monolayers},}\ }\href {\doibase 10.1073/pnas.1213930109}
  {\bibfield  {journal} {\bibinfo  {journal} {Proc. Natl. Acad. Sci. (USA)}\
  }\textbf {\bibinfo {volume} {109}},\ \bibinfo {pages} {16429--16433}
  (\bibinfo {year} {2012})}\BibitemShut {NoStop}%
\bibitem [{\citenamefont {Hasnain}\ \emph {et~al.}(2013)\citenamefont
  {Hasnain}, \citenamefont {Jungblut},\ and\ \citenamefont
  {Dellago}}]{Hasnain13}%
  \BibitemOpen
  \bibfield  {author} {\bibinfo {author} {\bibfnamefont {J.}~\bibnamefont
  {Hasnain}}, \bibinfo {author} {\bibfnamefont {S.}~\bibnamefont {Jungblut}}, \
  and\ \bibinfo {author} {\bibfnamefont {C.}~\bibnamefont {Dellago}},\
  }\bibfield  {title} {\enquote {\bibinfo {title} {Dynamic phases of colloidal
  monolayers sliding on commensurate substrates},}\ }\href {\doibase
  10.1039/c3sm50458a} {\bibfield  {journal} {\bibinfo  {journal} {Soft Matter}\
  }\textbf {\bibinfo {volume} {9}},\ \bibinfo {pages} {5867--5873} (\bibinfo
  {year} {2013})}\BibitemShut {NoStop}%
\bibitem [{\citenamefont {McDermott}\ \emph {et~al.}(2013)\citenamefont
  {McDermott}, \citenamefont {Amelang}, \citenamefont {Reichhardt},\ and\
  \citenamefont {Reichhardt}}]{McDermott13a}%
  \BibitemOpen
  \bibfield  {author} {\bibinfo {author} {\bibfnamefont {D.}~\bibnamefont
  {McDermott}}, \bibinfo {author} {\bibfnamefont {J.}~\bibnamefont {Amelang}},
  \bibinfo {author} {\bibfnamefont {C.~J.~Olson}\ \bibnamefont {Reichhardt}}, \
  and\ \bibinfo {author} {\bibfnamefont {C.}~\bibnamefont {Reichhardt}},\
  }\bibfield  {title} {\enquote {\bibinfo {title} {Dynamic regimes for driven
  colloidal particles on a periodic substrate at commensurate and
  incommensurate fillings},}\ }\href {\doibase 10.1103/PhysRevE.88.062301}
  {\bibfield  {journal} {\bibinfo  {journal} {Phys. Rev. E}\ }\textbf {\bibinfo
  {volume} {88}},\ \bibinfo {pages} {062301} (\bibinfo {year}
  {2013})}\BibitemShut {NoStop}%
\bibitem [{\citenamefont {Loehr}\ \emph {et~al.}(2016)\citenamefont {Loehr},
  \citenamefont {Loenne}, \citenamefont {Ernst}, \citenamefont {de~las Heras},\
  and\ \citenamefont {Fischer}}]{Loehr16}%
  \BibitemOpen
  \bibfield  {author} {\bibinfo {author} {\bibfnamefont {J.}~\bibnamefont
  {Loehr}}, \bibinfo {author} {\bibfnamefont {M.}~\bibnamefont {Loenne}},
  \bibinfo {author} {\bibfnamefont {A.}~\bibnamefont {Ernst}}, \bibinfo
  {author} {\bibfnamefont {D.}~\bibnamefont {de~las Heras}}, \ and\ \bibinfo
  {author} {\bibfnamefont {T.~M.}\ \bibnamefont {Fischer}},\ }\bibfield
  {title} {\enquote {\bibinfo {title} {Topological protection of multiparticle
  dissipative transport},}\ }\href {\doibase 10.1038/ncomms11745} {\bibfield
  {journal} {\bibinfo  {journal} {Nature Commun.}\ }\textbf {\bibinfo {volume}
  {7}},\ \bibinfo {pages} {11745} (\bibinfo {year} {2016})}\BibitemShut
  {NoStop}%
\bibitem [{\citenamefont {Tierno}\ \emph {et~al.}(2007)\citenamefont {Tierno},
  \citenamefont {Johansen},\ and\ \citenamefont {Fischer}}]{Tierno07}%
  \BibitemOpen
  \bibfield  {author} {\bibinfo {author} {\bibfnamefont {P.}~\bibnamefont
  {Tierno}}, \bibinfo {author} {\bibfnamefont {T.~H.}\ \bibnamefont
  {Johansen}}, \ and\ \bibinfo {author} {\bibfnamefont {T.~M.}\ \bibnamefont
  {Fischer}},\ }\bibfield  {title} {\enquote {\bibinfo {title} {Localized and
  delocalized motion of colloidal particles on a magnetic bubble lattice},}\
  }\href {\doibase 10.1103/PhysRevLett.99.038303} {\bibfield  {journal}
  {\bibinfo  {journal} {Phys. Rev. Lett.}\ }\textbf {\bibinfo {volume} {99}},\
  \bibinfo {pages} {038303} (\bibinfo {year} {2007})}\BibitemShut {NoStop}%
\bibitem [{\citenamefont {Grier}(2003)}]{Grier03}%
  \BibitemOpen
  \bibfield  {author} {\bibinfo {author} {\bibfnamefont {D.~G.}\ \bibnamefont
  {Grier}},\ }\bibfield  {title} {\enquote {\bibinfo {title} {A revolution in
  optical manipulation},}\ }\href {\doibase 10.1038/nature01935} {\bibfield
  {journal} {\bibinfo  {journal} {Nature (London)}\ }\textbf {\bibinfo {volume}
  {424}},\ \bibinfo {pages} {810--816} (\bibinfo {year} {2003})}\BibitemShut
  {NoStop}%
\bibitem [{\citenamefont {Lib\'al}\ \emph {et~al.}(2006)\citenamefont
  {Lib\'al}, \citenamefont {Reichhardt}, \citenamefont {Jank\'o},\ and\
  \citenamefont {Reichhardt}}]{Libal06}%
  \BibitemOpen
  \bibfield  {author} {\bibinfo {author} {\bibfnamefont {A.}~\bibnamefont
  {Lib\'al}}, \bibinfo {author} {\bibfnamefont {C.}~\bibnamefont {Reichhardt}},
  \bibinfo {author} {\bibfnamefont {B.}~\bibnamefont {Jank\'o}}, \ and\
  \bibinfo {author} {\bibfnamefont {C.~J.~Olson}\ \bibnamefont {Reichhardt}},\
  }\bibfield  {title} {\enquote {\bibinfo {title} {Dynamics, rectification, and
  fractionation for colloids on flashing substrates},}\ }\href {\doibase
  10.1103/PhysRevLett.96.188301} {\bibfield  {journal} {\bibinfo  {journal}
  {Phys. Rev. Lett.}\ }\textbf {\bibinfo {volume} {96}},\ \bibinfo {pages}
  {188301} (\bibinfo {year} {2006})}\BibitemShut {NoStop}%
\bibitem [{\citenamefont {Brazda}\ \emph {et~al.}(2017)\citenamefont {Brazda},
  \citenamefont {July},\ and\ \citenamefont {Bechinger}}]{Brazda17}%
  \BibitemOpen
  \bibfield  {author} {\bibinfo {author} {\bibfnamefont {T.}~\bibnamefont
  {Brazda}}, \bibinfo {author} {\bibfnamefont {C.}~\bibnamefont {July}}, \ and\
  \bibinfo {author} {\bibfnamefont {C.}~\bibnamefont {Bechinger}},\ }\bibfield
  {title} {\enquote {\bibinfo {title} {Experimental observation of
  {S}hapiro-steps in colloidal monolayers driven across time-dependent
  substrate potentials},}\ }\href {\doibase 10.1039/c7sm00393e} {\bibfield
  {journal} {\bibinfo  {journal} {Soft Matter}\ }\textbf {\bibinfo {volume}
  {13}},\ \bibinfo {pages} {4024--4028} (\bibinfo {year} {2017})}\BibitemShut
  {NoStop}%
\bibitem [{\citenamefont {B\"uchler}\ \emph {et~al.}(2003)\citenamefont
  {B\"uchler}, \citenamefont {Blatter},\ and\ \citenamefont
  {Zwerger}}]{Buchler03}%
  \BibitemOpen
  \bibfield  {author} {\bibinfo {author} {\bibfnamefont {H.~P.}\ \bibnamefont
  {B\"uchler}}, \bibinfo {author} {\bibfnamefont {G.}~\bibnamefont {Blatter}},
  \ and\ \bibinfo {author} {\bibfnamefont {W.}~\bibnamefont {Zwerger}},\
  }\bibfield  {title} {\enquote {\bibinfo {title} {Commensurate-incommensurate
  transition of cold atoms in an optical lattice},}\ }\href {\doibase
  10.1103/PhysRevLett.90.130401} {\bibfield  {journal} {\bibinfo  {journal}
  {Phys. Rev. Lett.}\ }\textbf {\bibinfo {volume} {90}},\ \bibinfo {pages}
  {130401} (\bibinfo {year} {2003})}\BibitemShut {NoStop}%
\bibitem [{\citenamefont {Muldoon}\ \emph {et~al.}(2012)\citenamefont
  {Muldoon}, \citenamefont {Brandt}, \citenamefont {Dong}, \citenamefont
  {Stuart}, \citenamefont {Brainis}, \citenamefont {Himsworth},\ and\
  \citenamefont {Kuhn}}]{Muldoon12}%
  \BibitemOpen
  \bibfield  {author} {\bibinfo {author} {\bibfnamefont {C.}~\bibnamefont
  {Muldoon}}, \bibinfo {author} {\bibfnamefont {L.}~\bibnamefont {Brandt}},
  \bibinfo {author} {\bibfnamefont {J.}~\bibnamefont {Dong}}, \bibinfo {author}
  {\bibfnamefont {D.}~\bibnamefont {Stuart}}, \bibinfo {author} {\bibfnamefont
  {E.}~\bibnamefont {Brainis}}, \bibinfo {author} {\bibfnamefont
  {M.}~\bibnamefont {Himsworth}}, \ and\ \bibinfo {author} {\bibfnamefont
  {A.}~\bibnamefont {Kuhn}},\ }\bibfield  {title} {\enquote {\bibinfo {title}
  {Control and manipulation of cold atoms in optical tweezers},}\ }\href
  {\doibase 10.1088/1367-2630/14/7/073051} {\bibfield  {journal} {\bibinfo
  {journal} {New J. Phys.}\ }\textbf {\bibinfo {volume} {14}},\ \bibinfo
  {pages} {073051} (\bibinfo {year} {2012})}\BibitemShut {NoStop}%
\bibitem [{\citenamefont {Schmidt}\ \emph {et~al.}(2018)\citenamefont
  {Schmidt}, \citenamefont {Lambrecht}, \citenamefont {Weckesser},
  \citenamefont {Debatin}, \citenamefont {Karpa},\ and\ \citenamefont
  {Schaetz}}]{Schmidt18}%
  \BibitemOpen
  \bibfield  {author} {\bibinfo {author} {\bibfnamefont {J.}~\bibnamefont
  {Schmidt}}, \bibinfo {author} {\bibfnamefont {A.}~\bibnamefont {Lambrecht}},
  \bibinfo {author} {\bibfnamefont {P.}~\bibnamefont {Weckesser}}, \bibinfo
  {author} {\bibfnamefont {M.}~\bibnamefont {Debatin}}, \bibinfo {author}
  {\bibfnamefont {L.}~\bibnamefont {Karpa}}, \ and\ \bibinfo {author}
  {\bibfnamefont {T.}~\bibnamefont {Schaetz}},\ }\bibfield  {title} {\enquote
  {\bibinfo {title} {Optical trapping of ion {C}oulomb crystals},}\ }\href
  {\doibase 10.1103/PhysRevX.8.021028} {\bibfield  {journal} {\bibinfo
  {journal} {Phys. Rev. X}\ }\textbf {\bibinfo {volume} {8}},\ \bibinfo {pages}
  {021028} (\bibinfo {year} {2018})}\BibitemShut {NoStop}%
\bibitem [{\citenamefont {Tung}\ \emph {et~al.}(2006)\citenamefont {Tung},
  \citenamefont {Schweikhard},\ and\ \citenamefont {Cornell}}]{Tung06}%
  \BibitemOpen
  \bibfield  {author} {\bibinfo {author} {\bibfnamefont {S.}~\bibnamefont
  {Tung}}, \bibinfo {author} {\bibfnamefont {V.}~\bibnamefont {Schweikhard}}, \
  and\ \bibinfo {author} {\bibfnamefont {E.~A.}\ \bibnamefont {Cornell}},\
  }\bibfield  {title} {\enquote {\bibinfo {title} {Observation of vortex
  pinning in {Bose-E}instein condensates},}\ }\href {\doibase
  10.1103/PhysRevLett.97.240402} {\bibfield  {journal} {\bibinfo  {journal}
  {Phys. Rev. Lett.}\ }\textbf {\bibinfo {volume} {97}},\ \bibinfo {pages}
  {240402} (\bibinfo {year} {2006})}\BibitemShut {NoStop}%
\bibitem [{\citenamefont {Veshchunov}\ \emph {et~al.}(2016)\citenamefont
  {Veshchunov}, \citenamefont {Magrini}, \citenamefont {Mironov}, \citenamefont
  {Godin}, \citenamefont {Trebbia}, \citenamefont {Buzdin}, \citenamefont
  {Tamarat},\ and\ \citenamefont {Lounis}}]{Veshchunov16}%
  \BibitemOpen
  \bibfield  {author} {\bibinfo {author} {\bibfnamefont {I.~S.}\ \bibnamefont
  {Veshchunov}}, \bibinfo {author} {\bibfnamefont {W.}~\bibnamefont {Magrini}},
  \bibinfo {author} {\bibfnamefont {S.~V.}\ \bibnamefont {Mironov}}, \bibinfo
  {author} {\bibfnamefont {A.~G.}\ \bibnamefont {Godin}}, \bibinfo {author}
  {\bibfnamefont {J.~B.}\ \bibnamefont {Trebbia}}, \bibinfo {author}
  {\bibfnamefont {A.~I.}\ \bibnamefont {Buzdin}}, \bibinfo {author}
  {\bibfnamefont {Ph.}\ \bibnamefont {Tamarat}}, \ and\ \bibinfo {author}
  {\bibfnamefont {B.}~\bibnamefont {Lounis}},\ }\bibfield  {title} {\enquote
  {\bibinfo {title} {Optical manipulation of single flux quanta},}\ }\href
  {\doibase 10.1038/ncomms12801} {\bibfield  {journal} {\bibinfo  {journal}
  {Nature Commun.}\ }\textbf {\bibinfo {volume} {7}},\ \bibinfo {pages} {12801}
  (\bibinfo {year} {2016})}\BibitemShut {NoStop}%
\bibitem [{\citenamefont {Lib\'al}\ \emph {et~al.}(2018)\citenamefont
  {Lib\'al}, \citenamefont {Nisoli}, \citenamefont {Reichhardt},\ and\
  \citenamefont {Reichhardt}}]{Libal18}%
  \BibitemOpen
  \bibfield  {author} {\bibinfo {author} {\bibfnamefont {A.}~\bibnamefont
  {Lib\'al}}, \bibinfo {author} {\bibfnamefont {C.}~\bibnamefont {Nisoli}},
  \bibinfo {author} {\bibfnamefont {C.~J.~O.}\ \bibnamefont {Reichhardt}}, \
  and\ \bibinfo {author} {\bibfnamefont {C.}~\bibnamefont {Reichhardt}},\
  }\bibfield  {title} {\enquote {\bibinfo {title} {Inner phases of colloidal
  hexagonal spin ice},}\ }\href {\doibase 10.1103/PhysRevLett.120.027204}
  {\bibfield  {journal} {\bibinfo  {journal} {Phys. Rev. Lett.}\ }\textbf
  {\bibinfo {volume} {120}},\ \bibinfo {pages} {027204} (\bibinfo {year}
  {2018})}\BibitemShut {NoStop}%
\bibitem [{\citenamefont {Nisoli}(2018)}]{Nisoli18}%
  \BibitemOpen
  \bibfield  {author} {\bibinfo {author} {\bibfnamefont {C.}~\bibnamefont
  {Nisoli}},\ }\bibfield  {title} {\enquote {\bibinfo {title} {Unexpected
  phenomenology in particle-based ice absent in magnetic spin ice},}\ }\href
  {\doibase 10.1103/PhysRevLett.120.167205} {\bibfield  {journal} {\bibinfo
  {journal} {Phys. Rev. Lett.}\ }\textbf {\bibinfo {volume} {120}},\ \bibinfo
  {pages} {167205} (\bibinfo {year} {2018})}\BibitemShut {NoStop}%
\bibitem [{\citenamefont {Ortiz-Ambriz}\ \emph {et~al.}()\citenamefont
  {Ortiz-Ambriz}, \citenamefont {Nisoli}, \citenamefont {Reichhardt},
  \citenamefont {Reichhardt.},\ and\ \citenamefont {Tierno}}]{OrtizAmbriz19}%
  \BibitemOpen
  \bibfield  {author} {\bibinfo {author} {\bibfnamefont {A.}~\bibnamefont
  {Ortiz-Ambriz}}, \bibinfo {author} {\bibfnamefont {C.}~\bibnamefont
  {Nisoli}}, \bibinfo {author} {\bibfnamefont {C.}~\bibnamefont {Reichhardt}},
  \bibinfo {author} {\bibfnamefont {C.~J.~O.}\ \bibnamefont {Reichhardt.}}, \
  and\ \bibinfo {author} {\bibfnamefont {P.}~\bibnamefont {Tierno}},\
  }\href@noop {} {\enquote {\bibinfo {title} {Ice rule and emergent frustration
  in particle ice and beyond},}\ }\bibinfo {note}
  {{arXiv:1909.13534}}\BibitemShut {NoStop}%
\bibitem [{M()}]{M}%
  \BibitemOpen
  \href@noop {} {}\bibinfo {note} {Supplementary movies are available at the
  following links: Dynamic chiral lattice at $F_{\rm trap}=0.4$,
  https://youtu.be/-39jaja\_b4c ; Choreographic lattice at $F_{\rm trap}=0.7$,
  https://youtu.be/YUWqG39CnJ0 ; Frustrated liquid state at $F_{\rm
  trap}=0.54$, https://youtu.be/9y\_8DoJdN\_E .}\BibitemShut {Stop}%
\end{thebibliography}%
\end{document}